\documentclass[aps,prl,twocolumn, showpacs,amsmath,amssymb,groupeaddress]{revtex4}
\usepackage{graphicx}
\usepackage{dcolumn}
\usepackage{bm}
\usepackage{amsmath}
\usepackage{amssymb}
\usepackage{epsfig}

\begin{document}
\title{Adiabatic Magnetization of Superconductors as a High-Performance Cooling Mechanism}
\author{Fabrizio Dolcini}
\email{f.dolcini@sns.it}
\affiliation{NEST CNR-INFM and Scuola Normale Superiore, I-56126 Pisa, Italy}
\author{Francesco Giazotto}
\email{f.giazotto@sns.it}
\affiliation{NEST CNR-INFM and Scuola Normale Superiore, I-56126 Pisa, Italy}
\begin{abstract}
The adiabatic magnetization of a superconductor is a cooling principle proposed in the 30s, which has never been exploited up to now. Here we present a detailed dynamic description of the effect,  computing the achievable  final temperatures as well as the process timescales for different  superconductors in various regimes. We show that, although in the experimental conditions explored so far the method is in fact inefficient, a suitable choice of initial temperatures and metals can lead to unexpectedly large cooling effect, even in the presence of dissipative phenomena. Our results suggest that this principle can be re-envisaged today as a performing refrigeration method to access the $\mu$K regime in nanodevices.
\\\\\\
\end{abstract}
\pacs{07.20.Mc,05.70.-a,74.25.Bt}

\maketitle
Since the very early discovery of the laws of thermodynamics, cooling represents one of the most fascinating challenges for both experimental and theoretical physics~\cite{pobell,rmp}. A well known cryogenic principle is the adiabatic demagnetization,  based on the property that ordinary magnetic materials, such as paramagnetic salts,   experience an entropy decrease   when  a magnetic field is applied, due to the alignment of their atomic dipoles.  This effect is currently applied also to nuclear spins under large magnetic fields,  allowing to reach the  $\mu$K regime in nuclear demagnetization refrigerators \cite{pobell}.  

\begin{figure}[t!] 
\epsfig{file=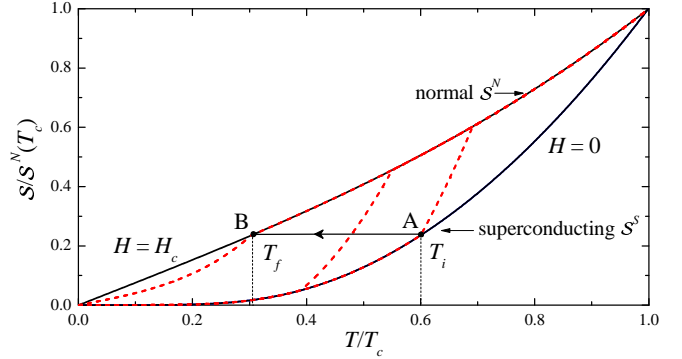,width=\columnwidth,clip=}
\caption{\label{Fig1}
(color online) Adiabatic magnetization of a superconductor. Solid curves describe the entropy $\mathcal{S}$ of a metal in the N and in the S state as a function of the reduced temperature. When a magnetic field is applied on a thermally-isolated superconductor at $T_i$, the metal is driven into the N state ($A \rightarrow B$), and the temperature decreases down to $T_f$. Dashed lines refer to the entropy in the intermediate state.}
\end{figure}

In superconducting materials, however, the opposite cooling principle is observed. 
It is well known~\cite{rickayzen} that a sufficiently strong magnetic field drives a superconductor~(S) into the normal~(N) state, and that such phase transition occurs with a supply of latent heat, since the S state is   much more ordered phase than N  at a given temperature. As a consequence, if a magnetic field with intensity $H$ increasing from 0 up to the critical value~$H_c$ is quasi-statically applied on a thermally isolated superconductor, its entropy $\mathcal{S}$ per unit volume is preserved 
\begin{equation} 
\mathcal{S}^{\text{N}}(T_f,H=H_c)=\mathcal{S}^{\text{S}}(T_i,H=0) \label{SE=SS},
\end{equation}
and the  metal cools from the initial temperature $T_i$ down  to a final temperature $T_f$ \cite{Shoenberg,rose-innes}, as illustrated  in Fig.~\ref{Fig1}. 
This cryogenic principle, known after the pioneering works by Mendelssohn \& Moore \cite{MM} and by Keesom \& Kok \cite{KK} in 1934 as "adiabatic magnetization of a superconductor" (AMS), offers the advantage that the required magnetic fields are much lower ($H \ll 1 {\rm T}$) than those typically used in adiabatic demagnetization refrigerators. Furthermore, the presence of a metal greatly simplifies the contacting to devices and allows faster equilibration time-scales. Although the validity of AMS was successively confirmed by other experiments~\cite{KvL,daunt,dolecek}, only a relatively small   cooling effect was observed so far:  the temperature was lowered  from $2.50$ K to $2.22$ K on tin samples \cite{MM,yaqub}, from $1.43$ K to $1.32$ K on thallium samples, and from $3.63$ K to $3.54$ K on lead spheres~\cite{dolecek}. It thus never became of practical use as a cryogenic technique, and its theoretical modeling has also been overlooked.  On the other hand, the exponential growth of nanotechnological applications at low temperature  demands  higher performance to refrigerators, which are required to be more versatile, faster and not invasive. The aforementioned features of AMS seem quite promising to this purpose,  and a detailed analysis of this refrigeration principle is desirable.

Here we present the first dynamical description of the adiabatic magnetization effect, taking into account the role of dissipative phenomena, and computing both the final temperature and the process time-scales. 
This analysis allows us to show that, while the conditions of the  experiments carried out so far were not suitable for cooling,  realistic regimes can be identified in which the adiabatic magnetization  can be exploited as a performing cooling technique.

We consider the case of type-I superconductors, and start our analysis with some remarks about thermodynamics.
In each phase the entropy includes a phonon and an electronic contribution, $\mathcal{S}^{\text{N}/\text{S}}=\mathcal{S}_{ph}+\mathcal{S}_{el}^{\text{N}/\text{S}}$. Explicitly, $\mathcal{S}_{ph} (T)=\alpha T^3$, where $T$ is the  temperature and $\alpha$ is the coefficient related to the Debye temperature.
The electronic entropy in the N state has a linear behavior $\mathcal{S}_{el}^{\text{N}}(T)=\gamma T$. 
The contribution of spin paramagnetism is negligible in the range of magnetic fields we are interested in, $H \ll 1 {\rm T}$, so that $\mathcal{S}^{\text{N}}(T,H_c)\cong \mathcal{S}^{\text{N}}(T,0)$. In the S phase   the condensate is a coherent state with vanishing entropy, so that $\mathcal{S}_{el}^{\text{S}}$ is purely due to quasi-particles and can be obtained from the BCS theory as
\begin{eqnarray}
\mathcal{S}_{el}^{\text{S}}(T)=-2\nu_{\text{F}}k_{\text{B}}\int^{\infty}_{-\infty} \! \! \! dE \, \mathcal{N}(E,T)    \left[f(E)\ln(f(E)) \right],
\label{Sentropy}
\end{eqnarray}
where $\nu_F$ is the normal density of states (DOS) at the Fermi level, $f(E)=(1+\exp[E/(k_{\text{B}}T)])^{-1}$ is the Fermi-Dirac  distribution function, $\mathcal{N}(E,T)=|E|/\sqrt{E^2-\Delta(T)^2}\Theta(E^2-\Delta(T)^2)$ is the BCS normalized DOS, with $\Delta(T)$ denoting the superconducting order parameter and $\Theta (x)$   the Heaviside   function.

For a given initial temperature $T_i$, the final temperature $T_f$  of the metal is determined by Eq.(\ref{SE=SS}), which can be rewritten as
\begin{equation}
 T_f + \frac{T^3_f}{(T^*)^2} \, = \, \frac{T^3_i}{(T^*)^2} \left[1+ \left( \frac{T^*}{T_c} \right)^2   \Phi\left(\frac{T_i}{T_c}\right) \right] , \label{TfTi}
\end{equation}
indicating that $T_f$ depends in general on {\it two} characteristic parameters, namely the critical temperature $T_c$ of the superconductor, and $T^* \doteq  \sqrt{\gamma/\alpha} = \sqrt{5 Z T^3_D/8 \pi^2 T_F}$,
which defines the temperature  below which the entropy of the N state is dominated by the electron contribution. 
Here, $Z$ denotes the nominal valence, while
$T_F$ and $T_D$  the Fermi and Debye temperatures of the metal, respectively.
Furthermore,  $\Phi$ is a universal function of $T/T_c$ defined through the relation $ \mathcal{S}_{el}^{\text{S}}(T)/\mathcal{S}_{ph}(T) = (T^*/T_c)^2 \, \Phi(T/T_c)$, exponentially small for $T/T_c < 0.1$, and of the order of unity for $0.5\leq T/T_c \leq1$. Despite the simplicity of its derivation, Eq. (\ref{TfTi}) contains important physical insight. Indeed if the initial temperature $T_i$ is of the same order as $T^*$, so is the final temperature $T_f$ ($T_f \lesssim T_i$), even if $T_i \ll T_c$. In this regime the AMS is therefore  clearly inefficient as a cooling mechanism. By contrast, if $T_i \ll T^*$, the final temperature decreases as  
\begin{equation}
T_f \simeq T^3_i/(T^*)^2.   \label{TfTi-approx}
\end{equation}
This \emph{cubic} dependence stems from the fact that, in this temperature regime, the AMS effectively transforms the entropy of phonons into the entropy of electrons. We emphasize that this effect represents an advantage with respect to the \emph{linear} gain factor $T_f/T_i$ characterizing the adiabatic demagnetization process \cite{pobell}.
\begin{figure} 
\epsfig{file=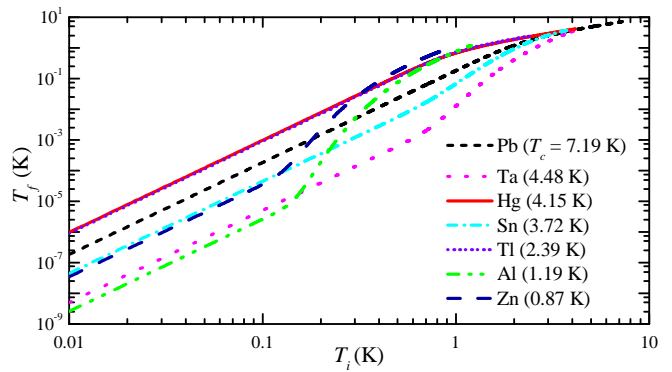,width=\columnwidth,clip=}
\caption{\label{Fig2} (color online)  Final  temperature $T_f$ vs initial temperature $T_i$  for several type-I superconductors, in the absence of dissipative effects.  }
\end{figure}
The efficiency of the AMS thus heavily depends on the material choice and on the initial temperature range.

Figure \ref{Fig2} shows $T_f$ vs $T_i$ calculated for several superconductors  from $10$ mK to the zero-field critical temperature $T_c$. The low-temperature linear behavior in the log-log plot accounts for the cubic dependence~(\ref{TfTi-approx}), and occurs in all materials. 
At higher temperatures, differences emerge between metals that exhibit $T^* < T_c$ (like Pb, Hg, Tl), and those with $T^* > T_c$ (like Al, Zn, and Ta). In the former case  $T_f \lesssim T_i$, whereas in the latter case $T_f$ exhibits a steep decrease governed by $T_c$, {\it i.e.} $T_f \simeq T_i^3 \, \Phi(T_i/T_c)/T_c^2$, before the crossover to the cubic law~(\ref{TfTi-approx}).  
It is noteworthy that most   experiments were concerned with the first group of metals, and with $T_i\sim T^*$. This  explains the unsatisfactory cooling observed on Sn \cite{MM,yaqub} and Pb \cite{dolecek}. Our analysis suggests that tantalum (Ta) is a good candidate, since $T^*/T_c \gtrsim 3$, and it allows to obtain $T_f$ in the range $\sim \mu \,{\rm K}...{\rm m}\,{\rm K}$ starting from $T_i$ in the range $\sim 100 {\rm m}\,{\rm K}...1\,{\rm K}$. Notice that Sn is suitable only if $T_i \leq 0.6 \,{\rm K}$, whereas Al is even better for $T_i \leq 0.2 \,{\rm K}$.

So far, using purely thermodynamical arguments and BCS theory, we have shown that AMS may in principle lead to extremely low values of $T_f$, provided the superconducting metal and initial temperature are appropriately chosen. However, the influence of  dissipative effects must be taken into account in order for such a method to  be considered as a promising cooling technique. 
First of all, when a magnetic field is applied to a type-I superconductor, the  transition to the N phase is preceded by formation of an intermediate state (IS), where S and N phases coexist  for $H^\prime_c(T)<H<H_c(T)$. 
Here  $H_c^\prime(T)=(1-n) H_c(T)$, $n$ is  the demagnetization factor, $H_c(T)$ is the critical field defined through the relation $dH_c^2(T)/dT=(2/\mu_0)[\mathcal{S}^{\text{S}}(T,0)-\mathcal{S}^{\text{N}}(T,0)]$ \cite{Shoenberg}, and $\mu_0$ is the vacuum permeability. 
The presence of a normal fraction $x_N$ in the IS yields dissipative eddy currents when the magnetic field is increased with time. 
Furthermore, in a cryostat the superconductor is connected to some mounting support that remains at the initial temperature $T_i$; the metal is thus exposed to a heat flux, which cannot be neglected in view of its relatively small low-temperature specific heat. Finally, once the cooling is realized,  heating generated by measurements on any device attached to the cryostat has to be considered. 
Thus, even assuming that the  range of initial temperatures and the superconductor are properly chosen, the existence of dissipative effects leads to the following questions: i) is the AMS-based cooling robust against these effects? ii)  if so, what are the typical time-scales in which low temperatures are reached, and how long can these be maintained? \\

To address these questions, we have analyzed the AMS {\it dynamically}, i.e., the process is governed by the equation
$\partial_t \mathcal{S} = P(t)/T(t)$, where $P$ is the total dissipated power per unit volume which involves the three contributions mentioned above. For simplicity, we consider a bundle of $N_w$ long and thin superconducting wires of radius $R$ and length $L$ each, attached to an insulating support of length $l$, as sketched in the inset of Fig. 3(a). The AMS is driven by the magnetic field, and three time regimes can be distinguished. In the first one the magnetic field is increased from 0 to $H^\prime_c(T_i)$, and no cooling occurs since the whole system remains  superconducting, so that $P=0$ and $T=T_i$. 
In the second regime (cool-down) $H$ is varied from $H^\prime_c(T_i)$ up to $H_c(0)$ over a time $\tau$, and the system enters into the IS state. 
Although a detailed description of the IS  for a given geometry is, in general,  quite complicated, we wish here to capture its main characteristics. We shall thus follow Ref. \onlinecite{rose-innes} by assuming that the N and S regions of the IS are uniformly distributed, so that the normal fraction $x_N$ increases with the magnetic field as $x_N(T,H)= 1-n^{-1} \left( 1-H/H_c(T)\right)$, and that the entropy is a linear combination of the N and S entropies, $\mathcal{S}(T,H)= x_N \,   \mathcal{S}^{\rm N}(T,0)  \, + (1-x_N)  \mathcal{S}^{\rm S}(T,0)$. In this case the AMS is described by the differential equation  
\begin{equation}
\mathcal{C}_V(T,H) \, \dot{T} -\frac{\mu_0}{n} \,T \, \frac{dH_c(T)}{dT}\, \dot{H} =P \quad,
\end{equation}
where 
\begin{equation}
\mathcal{C}_V(T,H) = x_N \, \mathcal{C}_V^{\rm N}(T) \, + (1-x_N) \, \mathcal{C}_V^{\rm S}(T) \, + \mathcal{C}_V^{lat}(T,H) 
\label{totalC}
\end{equation} 
is the  total specific heat (per unit volume)  in the IS,
$\mathcal{C}_V^{\rm N/S}(T)=T \, \partial \mathcal{S}^{\rm{S/N}}(T)/\partial T$, and $\mathcal{C}_V^{lat}=(TH/\mu_0nH_c^3(T))(\mathcal{S}^{\text{N}}(T,0)-\mathcal{S}^{\text{S}}(T,0))^2$ is  the latent heat to be supplied for the transition.
As soon as $x_N \neq 0$ the cooling mechanism is enabled, and the temperature of the metal starts to lower, though contrasted by the dissipative effects. 
A simple calculation shows that the variation of the magnetic field $B=x_N \mu_0 H_c(T)$ trapped in the normal fraction  induces  eddy current dissipation  $P_{eddy}(t)=R^2 \sigma \dot{B}^2/8$ in each wire, where $\sigma$ is the electric conductivity of the metal. At the same time, the temperature gradient  across the insulating support  between the 'hot' upper surface at temperature $T_i$ and the 'cold' lower surface in contact with the metal [see the inset of Fig. 3(a)] leads to a heat flow $P_{supp} = b (T_i^{\beta+1}-T^{\beta+1})/ (\beta+1) l L$. Here, $b$ and $\beta$ are  parameters characterizing the temperature dependence of thermal conductivity $\kappa_{supp}(T)=b T^\beta$ of the insulating support~\cite{pobell}.  Any temperature gradient in the metal has been neglected due to its relatively high thermal conductivity.
\begin{figure} 
\epsfig{file=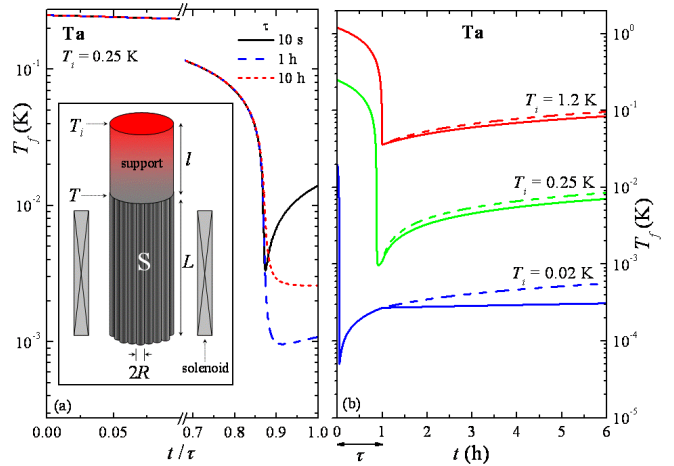,width=\columnwidth,clip=}
\caption{\label{Fig3} (color online) (a) Time evolution of $T_f$ in the intermediate state calculated at $T_i=250$ mK for three values of $\tau$. The inset shows a scheme of the adiabatic magnetization cooler.
(b)  Full time evolution of $T_f$ calculated at three different $T_i$ for $\tau=1$ h. Dashed curves refer to $P_{load}\neq0$: from bottom to top, $P_{load}=10$ pW, 1 nW, and 100 nW. All calculations were performed for Ta. (see text)
}
\end{figure}

Figure~3(a) displays the temperature evolution for $0 \le t \le \tau$ calculated for three values of $\tau$, starting from $T_i=0.25$ K.  
For simplicity we assumed $H$ to vary linearly with time.
For  $t \ll \tau$ the temperature experiences a relative slow decrease, whereas a fast drop is observed for $t \lesssim \tau$. 
The cooling effect is eventually contrasted by  both dissipative eddy currents and heat flow from the support. 
It is worth emphasizing that $P_{eddy}$ and $P_{supp}$  behave differently with respect to the velocity of the magnetic field variation. 
For fast field variations [solid curve of Fig.~3(a)] $P_{eddy}$   is relevant and $P_{supp}$ is suppressed. 
In contrast, for slow field variations [dotted curve of Fig. 3(a)] Joule heating  has a minor effect while the heat flow from support affects the cooling for longer time. For a given geometry and initial temperature,  the competition of these two terms  determines the optimal time which allows to reach the lowest temperature, as shown by the dashed curve in Fig.~3(a) for $\tau=1$ h. Such time-scale depends on the electric conductivity of the metal, and the thermal conductivity of the support. 
Tantalum seems to be a good candidate superconductor due to its  relatively low conductivity $\sigma \sim 10^9 \,\Omega^{-1} {\rm m}^{-1}$, and  high specific heat ($\gamma \sim 523 \,{\rm J m^{-3} K^{-2}}$ and $\alpha \sim 2.63 \,{\rm J m^{-3} K^{-4}}$). We note that aluminum (Al), in spite of its high ratio $T^*/T_c$, is   less suitable for AMS due to its extremely high electric conductivity; alternatively, tin (Sn) may be a fair choice. As far as the  support is concerned, a good insulator like PVC, with parameters $b=1.8\times10^{-5}$ Wm$^{-1}$K$^{-1}$ and $\beta=2.05$~\cite{pobell}, seems appropriate. The plots of Fig. 3 refer to this case~\cite{parameters}.

The last time-regime corresponds to the case where the system is fully normal. The magnetic field is not further varied   [$H \equiv H_c(0)$], so that $P_{eddy}=0$. This is the regime where measurements are typically carried out on a device thermally anchored to the metal. We have thus included a constant load power for $t>\tau$ arising from the measurement ($P_{load}$)   besides $P_{supp}$. The result of the whole dynamical process is shown in Fig. \ref{Fig3}(b) for three initial temperatures $T_i$, corresponding to the base temperature of a $^4{\rm He}$ cryostat ($T_i=1.2$ K), a $^3{\rm He}$ cryostat ($T_i=250$ mK), and a dilution refrigerator ($T_i=20$ mK).  For each $T_i$, the solid (dashed)  curve represents the temperature evolution  without (with) $P_{load}$. 
Notably, this cooling method ensures in all these ranges of operation a temperature gain of about \emph{two} orders of magnitude, which can be reached within an hour or less, and can be maintained for several hours. 
This represents an advantage with respect to the time-scales typical of the adiabatic demagnetization of nuclei.
The external load sustained by the AMS method depends  on the temperature range of operation. 
For instance, we have calculated that a bundle of wires of about $4 \times 10^3 \,{\rm cm}^3$ volume operating at $T_i=1.2$ K  can sustain   a power load of $100$ nW without significantly affecting its final temperature, whereas at $T_i=20$ mK a power load of $10$ pW increases $T_f$ of few hundreds of $\mu$K after 5 hours of operation [see Fig. 3(b)]. We stress that ordinary superconducting electronics (such as tunnel junctions circuits, radiation detectors, and SQUIDs) as well as single-electron devices exhibit power dissipation typically below 1 ${\rm pW}$, thus suggesting that AMS is   suitable to operate on nanostructures in the ultra-low temperature regime. 

Finally we notice that, since the magnetization  must  evolve through equilibrium states, the   variation of the applied magnetic field must proceed slow enough for relaxation processes to ensure equilibrium between electrons and lattice phonons. 
The determination of such time-scales in the IS is a crucial issue, since  the N and S phases have much different characteristic relaxation rates \cite{kaplan}. Analyzing the three terms of Eq.~(\ref{totalC}), one can easily prove that
  even a small normal fraction $x_N \sim 10^{-3}$  is sufficient for $\mathcal{C}_V^{\rm N}$ to largely dominate the other two contributions. 
Thus, apart from an extremely small range of magnetic fields, the specific heat of the superconductor in the IS is essentially determined by the electronic contribution in the  N fraction, which drives the cooling "dragging" the lattice phonons and the S fraction. 
An upper bound for the relaxation time characterizing the process is therefore represented by the inverse of the electron-phonon scattering rate in the N phase, which scales as $\tau^{-1}_{el-ph} \propto   T^{3}$ \cite{rmp}, and is typically much shorter  than that of the superconducting phase. For Ta   in the temperature range $2\times 10^{-4}\ldots 5\times 10^{-1}$ K, $\tau_{el-ph}$ lies in the range $\sim   10^{-7} \ldots 10^{3}$ s \cite{kaplan}, thus ensuring the consistency of our quasi-static approach.

In conclusion, we have provided a dynamic description of cooling by  adiabatic magnetization of superconductors. 
We have shown that, while in the experimental conditions  explored   so far the method is in fact inefficient, a suitable choice of temperature ranges and superconductors make this principle promising as a high-performance  refrigeration technique. Beside involving low magnetic fields (i.e., $\sim 10^{-2}$ to $10^{-1}$ {\rm T}), the present method offers the additional advantage that the final temperature  depends cubically on the initial one [see Eq.(\ref{TfTi-approx})]. Moreover, we find that the cool-down times  are comparable or shorter than those of typical demagnetization cryostats in the same temperature range, while the warming-up rates can be of the order of several hours under continuous power load (see Fig. \ref{Fig3}). Our results suggest that  magnetization cycles to improve this cooling principle can also be envisioned.

We acknowledge  R.~Fazio and J.~P.~Pekola  for fruitful discussions, and the NanoSciERA "NanoFridge" EU project and "Rientro dei Cervelli" MIUR program  for financial support.

\end{document}